\documentclass[manuscript]{aastex}

\begin{document}
\pagenumbering{arabic}

\title{EVOLUTION OF GALAXY MORPHOLOGY}
\author{Sidney van den Bergh}
\affil{Dominion Astrophysical Observatory, Herzberg Institute of Astrophysics, National Research Council of Canada, 5071 West Saanich Road, Victoria, BC, Canada  V9E 2E7}
\email{sidney.vandenbergh@nrc.ca}

\begin{abstract}

A strictly empirical review is given of presently available data 
on the evolution of galaxy morphology. From HST observations of 
distant galaxies and ground-based observations of nearby ones observed 
at the same rest-frame wavelength it is found that late-type (Sbc-Sc) 
galaxies evolve more rapidly with increasing redshift than do early-type
(E-Sa-Sab) galaxies. Furthermore the fraction of peculiar objects, that
cannot be shoehorned into the Hubble tuning fork classification scheme,
increases rapidly with redshift. Unexpectedly it is found that, over a
wide range of densities, the fraction of barred galaxies is independent
of environment. However, this fraction of barred galaxies appears to
decline rapidly with increasing redshift. 

\begin{center}
{\it Don't assume anything - Look!}\\
Gen. Joe Stilwell\\
\end{center}

\end{abstract}


\section{INTRODUCTION}

   Theories of galaxy evolution remain speculative and uncertain. 
However, strong constraints on such theories are becoming available
from the marvelous new imaging  data on galaxy morphology at various 
redshifts that have have become available during the last seven years. 
Such observations show that most star formation in galaxies with $z < 1$ 
takes place in disks, whereas stars in objects with $z > 2$ occurs mainly 
in luminous ``blobs'' and chaotic structures. Furthermore the typical 
galaxy at $z >2$ is, at any given time, undergoing a merger, whereas such 
major mergers are seen to be relatively rare at $z < 1$. Additionally,
late-type galaxies are seen to change their appearance rapidly with 
increasing redshift, whereas the morphological evolution of early-type 
galaxies seems to be much slower. Unexpectedly the fraction of barred 
galaxies is observed to be a steeply decreasing function of redshift, 
while the fraction of nearby barred galaxies is found to be almost 
independent of their environment.

\section{BARRED GALAXIES}
   Some time ago van den Bergh et al. (1996) noticed that barred
galaxies appeared to be much rarer in the Hubble Deep Field than they
are in nearby regions of the Universe. Making detailed corrections for
band-shift effects, changes in resolution, and the increase of noise in
observations of more distant galaxies, van den Bergh et al. (2002) were
able to confirm the reality of this effect. As viewed in rest frame blue
light, the fraction of barred galaxies appears to decrease from 23\% at 
$z = 0.0$ to $\sim4\%$ at $z \sim 0.7$. Possibly this decrease in the fraction of 
barred galaxies with increasing redshift is due to the fact that young,
recently formed, spiral galaxies are still too chaotic (dynamically
``hot'') to undergo global bar-like instabilities. Another unexpected 
effect [see Table 1] is that the frequency of of bars in disk (S0-Im)
galaxies appears to be almost independent of galaxy environment (van den Bergh 2002, and work in preparation).

\begin{deluxetable}{lc}
\tablewidth{0pt}

\tablecaption{DEPENDENCE OF FRACTION OF BARRED GALAXIES ON ENVIRONMENT}

\tablehead {\colhead{Enviroment} & \colhead{Percentage of barred galaxies}}

\startdata

Nearby field       &     $25 \pm 3$     \\
Nearby groups      &     $19 \pm 4$     \\
Nearby clusters    &     $28 \pm 3$      \\
Coma               &     $32 \pm 5$      \\

\enddata
\end{deluxetable}

   Among 1103  nearby disk galaxies that have been observed with 
large reflecting telescopes (Sandage \& Tammann 1981), it is found 
that 26 $\pm$ 2\% are barred. This value does not appear to differ 
significantly from 32 $\pm$ 5\% barred objects among 107 Coma S0-Sc 
galaxies with $m < 17.0$.
   Since the fraction of barred objects depends slightly on Hubble 
type it is, perhaps, fairest to compare the fraction of barred 
galaxies of types S0 + S0/a in the entire Shapley-Ames catalog directly with the corresponding fraction in the Coma cluster. For 
the entire Shapley-Ames catalog  25 $\pm$ 4\% of 190 S0 + S0/a 
galaxies are barred, compared to  24 $\pm$ 6\% of 62.5 such objects 
among galaxies with $m < 17.0$ in the Coma cluster. This result 
suggests that that the process that results in the formation of 
galactic bars is an internal one that is almost independent of 
galaxy environment.

\section{PECULIAR GALAXIES}

   A galaxy is defined as being ``peculiar'' if it differs in some
significant way from the prototypes used to define the Hubble 
tuning fork classification system. It is one of the beauties of 
Hubble's system that most luminous nearby galaxies fit it so 
well, and do not need to be "shoehorned" into the system. One
of the most striking results obtained from the imaging of the
Hubble Deep Field (Ferguson, Dickinson \& Williams 2000) was
that such a large fraction of the HDF(N) galaxies had peculiar
morphologies. From my own classifications I find that the overall
fraction of peculiar galaxies (as viewed in the restframe blue band)
increases from 12\% at z = 0.0 to 46\% at $z \sim 0.7$. However, these
overall figures are a bit misleading because of the difference
that is observed between the way in which early-type and late-type
galaxies ``age''. At $z \sim 0.7$ only $\sim$5\% of E-S0-Sa galaxies appear
to be peculiar. For comparison 69\% of Sbc-Sc galaxies are 
peculiar at $z \sim 0.7$. Taken at face value this result suggests
that late-type galaxies have required a longer time arrive
at their present morphology than have systems of early type. Also 
the nature of the peculiarities seen in early-type galaxies are systematically different from those that are observed in
late-type systems. In distant Sa-Sb spirals the arm structure 
is generally less well-developed than it is among nearby spirals
with $z \sim 0$. On the other hand the peculiarity of late-type
spirals at high redshifts is mainly due to the fact that their 
spiral structure is more chaotic than it is for nearby Sbc and
Sc galaxies. A special kind of peculiarity occurs among spirals 
that are located in dense cluster environments (van den Bergh 1976).
As a result of what is nowadays referred to as ``galaxy harassment''
(Moore et al. 1996) the spiral arms of tidally interacting
early-type galaxies (and galaxies in rich clusters) have a ``fuzzy''  
appearance. [In extreme cases the spiral structure becomes ``anemic''.]
On the other hand the spiral arms of late-type interacting
spirals take on a ``knotty'' morphology, which is  presumably due to 
an enhanced formation rate of clusters and associations.  

\section{THE MADAU PLOT}
   Perhaps the most striking feature of galaxy morphology and
star formation is that most stars at $z < 1$ appear to be
forming in disks. On the other hand the majority of stars in
objects with $z >2$ seem to form in luminous clumps or in chaotic 
structures. Possibly the change in slope of the Madau (1997)
plot near $z \sim 1.5$ is due to this transition from chaotic/clumpy
star formation at high redshifts to star creation in disks 
among the majority of nearby galaxies.

\section{MERGERS}
   On deep exposures with the HST the surface density of galaxies
is high. As a result many galaxies are members of optical, i.e.
nonphysical pairs. One can try to circumvent this problem
by only accepting those objects which either (1) exhibit
(tidal) distortions, or (2) galaxies with physically overlapping 
main bodies as merger candidates. Adopting this definition it
is found that only $\sim$5\% of the galaxies in the HDF(N) +HDF(S) that
have $z < 1.2$ are merger candidates. On the other hand it turns
out that $\sim$57\% of the objects with $z > 2$.0 are merger candidates.
In other words most galaxies at $z > 2$ are, at any given time,
involved in a merger with a luminous (massive) companion, whereas
nearby galaxies are typically single. It is noted in passing that 
only 1.5\% of the galaxies with $m < 17.0$ in the Coma cluster appear 
to be merging (or have double or multiple cores). Presumably this 
low merger frequency is a direct consequence of the very high
(1038 $\pm$ 60 km/s, Colless \& Dunn 1996) velocity dispersion in 
the Coma cluster. 

\section{CONCLUSIONS}
   Perhaps the most important insight that has been obtained
in recent times is that galaxy morphology depends strongly on 
both the environment and on the redshifts of galaxies. The 
Hubble tuning fork classification system is strictly applicable
only to nearby galaxies with $z < 0.5$, with bars apparently
becoming ever less frequent with increasing redshift. Furthermore 
The Hubble system becomes degenerate in very dense environments 
where the majority of galaxies are of Hubble types E, S0 and SB0.

\end{document}